\documentclass{article}

\title {Rutherford scattering with radiation damping}

\author{C.E. Aguiar and F.A. Barone\footnote{Present address:
Centro Brasileiro de Pesquisas F{\'\i}sicas, Rio de Janeiro, Brasil}
\\ \\ 
Instituto de F{\'\i}sica, Universidade Federal do Rio de Janeiro, Brasil}

\date{}

\usepackage{graphicx}
 
\begin {document}

\maketitle

\begin{abstract}
We study the effect of radiation damping on the classical scattering of 
charged particles.
Using a perturbation method based on the Runge-Lenz vector, we 
calculate radiative corrections to the Rutherford cross section, 
and the corresponding energy and angular momentum losses.
\end{abstract}


\section{Introduction}

The reaction of a classical point charge to its own radiation was
first discussed by Lorentz and Abraham more than one hundred years 
ago, and never stopped being a source of controversy and fascination 
\cite{RohrlichBook,Landau,Jackson,Rohrlich}. 
Nowadays, it is probably fair to say that the most disputable 
aspects of the Lorentz-Abraham-Dirac theory, like runaway solutions 
and preacceleration, have been adequately understood and treated 
in terms of finite-size effects (for a review see Ref.~\cite{Rohrlich}).
In any case, radiation damping considerably complicates the equations 
of motion of charged particles, and for many basic problems, like 
Rutherford scattering, only numerical calculations of the trajectories 
are available \cite{Plass,Huschilt}.
In this paper we study the effect of radiation reaction on the 
classical two-body scattering of charged particles. 
Following Landau and Lifshitz \cite{Landau}, we expand the 
electromagnetic force in powers of $c^{-1}$ ($c$ is the speed of light), 
up to the order $c^{-3}$ where radiation damping appears.
Then, using a perturbation technique based on the Runge-Lenz vector 
\cite{Aguiar}, we calculate the radiation damping corrections 
to the Rutherford deflection function and scattering cross section, and 
the corresponding expressions for the angular momentum and energy losses.

This paper is organized as follows. In Sec.~2 we obtain the radiation
damping force on a system of charged particles, from the expansion of 
the electromagnetic field in powers of $1/c$. 
The equations of motion for a two-body system with radiation reaction 
are discussed in Sec.~3, and in Sec.~4 we use the Runge-Lenz 
vector to calculate the radiation effect on classical Rutherford 
scattering. Some final remarks are made in Sec.~5.


\section{The radiation damping force}

In this section we reproduce, for completeness, the derivation of the
radiation damping force given by Landau and Lifshitz \cite {Landau}. 
We start from the electromagnetic potentials $\phi({\bf r},t)$ and 
${\bf\cal A}({\bf r},t)$, created by the charge and current densities 
$\rho({\bf r},t)$ and ${\bf J}({\bf r},t)$,
\begin {equation}
\label {scalar-pot}
\phi ({\bf r},t)=\int {\rho ({\bf r}',t_{R})\over R} d^{3}{\bf r}' \,,
\end {equation}
\begin {equation}
\label {vector-pot}
{\bf\cal A}({\bf r},t)=
  \frac{1}{c}\int {{\bf J}({\bf r}',t_{R})\over R} d^{3}{\bf r}' \,.
\end {equation}
Here, $R=|{\bf r}-{\bf r}'|$ and $t_{R}=t-R/c$ is the retarded time.
The electric and magnetic fields, ${\bf E}$ and ${\bf B}$, are obtained from 
the potentials as
\begin {equation}
{\bf E}=-\nabla\phi({\bf r},t)-\frac{1}{c}{\partial{\bf\cal A}({\bf r},t)\over\partial t}
\ \ \ ,\ \ \ 
{\bf B}=\nabla \times {\bf\cal A}({\bf r},t)\ .
\end {equation}
We want to calculate the electromagnetic force on a charge $q$,
\begin {equation}
{\bf F}=q{\bf E}+{q\over c}{\bf v}\times{\bf B} ,
\end {equation}
as a series in powers of $1/c$.
In order to do this, we expand $\rho ({\bf r}',t_{R})$ and 
${\bf J} ({\bf r}',t_{R})$ in Taylor series around $t_{R}=t$, 
\begin {eqnarray}
\label {rhoexpandido}
\rho({\bf r}',t_{R}) 
&=& \rho({\bf r}',t) +
   \frac{\partial\rho({\bf r}',t)}{\partial t}\left(-\frac{R}{c}\right) + 
   \frac{1}{2} \frac{\partial^{2}\rho({\bf r}',t)}{\partial t^{2}}
   \left(-\frac{R}{c}\right)^{2}
\cr &&
   +\frac{1}{6} \frac{\partial^{3}\rho({\bf r}',t)}{\partial t^{3}}
    \left(-\frac{R}{c}\right)^{3} + {\cal O}(c^{-4}) \,, 
\\
\label {Jexpandido}
{\bf J}({\bf r}',t_{R}) &=& {\bf J}({\bf r}',t) + 
   \frac{\partial{\bf J}({\bf r}',t)}{\partial t} \left(-\frac{R}{c}\right) + 
   {\cal O}(c^{-2}) \, .
\end {eqnarray}
Substituting these expansions in Eqs.(\ref {scalar-pot}) and (\ref {vector-pot}), 
and noting the charge conservation relation,
\begin {equation}
\frac{\partial}{\partial t} \int\rho({\bf r}',t) d^{3}{\bf r}' = 0 \,,
\end {equation}
we obtain
\begin {eqnarray}
\label {phiint}
\phi ({\bf r},t)
&=& \int \frac{\rho ({\bf r}',t)}{R} d^{3}{\bf r}' 
    + \frac{1}{2c^2}\frac{\partial^{2}}{\partial t^{2}}
      \int R\rho ({\bf r}',t)d^{3}{\bf r}'
\cr &&
    - \frac{1}{6c^3}\frac{\partial^{3}}{\partial t^{3}}
      \int R^{2}\rho ({\bf r}',t)d^{3}{\bf r}'
    +{\cal O}(c^{-4})
\\
\label {Aint}
\frac{1}{c}{\bf\cal A}({\bf r},t)
&=& \frac{1}{c^2}\int{{\bf J}({\bf r}',t)\over R}d^{3}{\bf r}'
    - \frac{1}{c^3}{\partial\over\partial t}\int{\bf J}({\bf r}',t)d^{3}{\bf r}'
    + {\cal O}(c^{-4}) \,.
\end {eqnarray}

With the gauge transformation
\begin {eqnarray}
\phi({\bf r},t)&\rightarrow&\phi({\bf r},t)-\frac{1}{c}{\partial\chi({\bf r},t)\over\partial t}\cr\cr
{\bf\cal A}({\bf r},t)&\rightarrow&{\bf\cal A}({\bf r},t)+{\bf\nabla}\chi({\bf r},t)\ ,
\end {eqnarray}
where
\begin {equation}
\chi({\bf r},t)={1\over 2c}{\partial\over\partial t}\int R\rho({\bf r},t)\ d^{3}{\bf r}'-{1\over 6c^{2}}{\partial^{2}\over\partial t^{2}}\int R^{2}\rho({\bf r},t)\ d^{3}{\bf r}'\ ,
\end {equation}
we can rewrite Eqs.~(\ref {phiint}) and (\ref {Aint}) as
\begin {eqnarray}
\phi ({\bf r},t) 
&=& \int {\rho ({\bf r}',t)\over R} d^{3}{\bf r}' + {\cal O}(c^{-4})
\\
\frac{1}{c}{\bf\cal A}({\bf r},t)
   &=&\frac{1}{c^2}\int {{\bf J}({\bf r}',t)\over R} d^{3}{\bf r}'
      + \frac{1}{2c^2}{\partial\over\partial t} 
          \int \frac{\bf R}{R} \rho({\bf r}',t) d^{3}{\bf r}'
\cr &&
      - \frac{1}{c^3}{\partial\over\partial t}
          \int{\bf J} ({\bf r}',t)d^{3}{\bf r}'
      - \frac{1}{3c^3}{\partial^{2}\over\partial t^{2}}
          \int{\bf R}\rho ({\bf r}',t) d^{3}{\bf r}'+ {\cal O}(c^{-4}) \,.
\end {eqnarray}
For a set of point charges $q_{k}$, with positions ${\bf r}_{k}(t)$ and 
velocities ${\bf v}_{k}(t)$, we have 
\begin {eqnarray}
\rho ({\bf r},t)&=&\sum_{k}q_{k}\delta\Bigl({\bf r}-{\bf r}_{k}(t)\Bigr) \,,
\\
{\bf J}({\bf r},t)&=&\sum_{k}q_{k}{\bf v}_{k}(t)\delta\Bigl({\bf r}-{\bf r}_{k}(t)\Bigr)\,,
\end {eqnarray}
and the potentials become 
\begin {eqnarray}
\label {potenciaisdiscretos1A}
\phi ({\bf r},t)&=&\sum_{k}{q_{k}\over R_{k}(t)} + {\cal O}(c^{-4})
\\
\frac{1}{c} {\bf\cal A}({\bf r},t)
&=&\frac{1}{c^2}\sum_{k}{q_{k}{\bf v}_{k}(t)\over R_{k}(t)}
   + {1\over 2c^2}{d\over dt}\sum_{k}{{\bf R}_{k}(t)\over R_{k}(t)}q_{k}
\cr&&
   - {1\over c^{3}}{d\over dt}\sum_{k}{q_{k}{\bf v}_{k}(t)}
   - {1\over 3c^{3}}{d^{2}\over dt ^{2}}\sum_{k}{\bf R}_{k}(t)q_{k}
   + {\cal O}(c^{-4})\, ,
\label {potenciaisdiscretos1B}
\end {eqnarray}
with ${\bf R}_{k}(t)={\bf r}-{\bf r}_{k}(t)$.
Carrying out the time derivatives in Eq.~(\ref{potenciaisdiscretos1B}) we obtain
\begin {eqnarray}
\label {Acorrigido}
\frac{1}{c}{\bf \cal A}({\bf r},t) 
&=& {1\over 2c^2}\sum_{k}\left[{q_{k}{\bf v}_{k}(t)\over R_{k}(t)}
    + {q_{k} {\bf R}_{k}(t).{\bf v}_{k}(t)\over R_{k}^{3}(t)}{\bf R}_{k}(t)\right] 
\cr && 
    - {2\over 3c^3}\sum_{k}q_{k}{\bf a}_{k}(t) + {\cal O}(c^{-4}) \,,
\end {eqnarray}
where ${\bf a}_{k}(t)$ is the acceleration of particle $k$.

The $\phi$ potential given in Eq.~(\ref {potenciaisdiscretos1A}) 
accounts for the Coulomb interaction.
The first term in Eq.~(\ref {Acorrigido}), of order $1/c^2$, introduces 
magnetic and retardation effects, and can be used to set up the Darwin 
lagrangian \cite{Landau}.
The last term in Eq.~(\ref {Acorrigido}), of order $1/c^3$, gives the radiation
damping electric field
\begin {equation}
\label {Erd}
{\bf E}_{rd}={2\over 3c^{3}}\sum_{k}q_{k}\frac{d{\bf a}_{k}}{dt} \,,
\end {equation}
and a null magnetic field ($\bf\cal A$ is independent of $\bf r$ in this order). 
Introducing the electric dipole of the system, 
${\bf D}=\sum_{k}q_{k}{\bf r}_{k}$, 
the radiation damping field of Eq.~(\ref{Erd}) can 
be written as
\begin {equation}
{\bf E}_{rd}= \frac{2}{3c^3}\frac{d^3{\bf D}}{dt^3}\, ,
\end {equation}
showing that it represents the reaction to the
electric dipole radiation emitted by the whole system.

The radiation damping force on charge $q_i$ is then
\begin {equation}
{\bf F}^{(i)}_{rd}={q_i\bf E}_{rd}={2\over 3c^{3}}\sum_{k}q_i q_{k}\frac{d{\bf a}_{k}}{dt}\,.
\end {equation}
It should be stressed that radiation reaction is not just
a self-force --- it gets contributions from every particle in the system. 
Only for a single accelerating charge $q$ the radiation damping force 
reduces to the Abraham-Lorentz self-interaction
\begin {equation}
\label {freiamento1part}
{\bf F}_{rd}={2\over 3}{q^{2}\over c^{3}}\frac{d{\bf a}}{dt}\ .
\end {equation}


\section {Two-body motion with radiation damping}

Let us consider a system of two charged particles. 
Taking radiation damping into account, their equations of motion read
\begin {eqnarray}
\label{eqmov-p1}
{d^2{\bf r}_{1}\over dt^2}&=& \frac{q_1q_2}{m_1}\frac{\bf r}{r^3}
+{2\over 3c^{3}}{q_{1}\over m_{1}}\frac{d}{dt}(q_{1}{\bf{a}}_{1}+q_{2}{\bf{a}}_{2})\,,
\\
\label {eqmov-p2}
{d^2{\bf r}_{2}\over dt^2}&=& -\frac{q_1q_2}{m_2}\frac{\bf r}{r^3}
+{2\over 3c^{3}}{q_{2}\over m_{2}}\frac{d}{dt}(q_{1}{\bf{a}}_{1}+q_{2}{\bf{a}}_{2})\,.
\end {eqnarray}
where ${\bf r}={\bf r}_1-{\bf r}_2$ and $m_i$ is the mass of particle $i$.
In these equations we have discarded the $c^{-2}$ terms that account for
the variation of mass with velocity and the Darwin magnetic and retardation 
effects. These terms do not interfere with our treatment of radiation damping,
and their effect on Rutherford scattering is discussed in Refs.~\cite{Aguiar,Aguiar2}. 

Subtracting Eq.~(\ref{eqmov-p2}) from (\ref{eqmov-p1}) 
we find
\begin {equation}
\label{eqmovrel1}
{d^{2}{\bf r}\over dt^{2}}=
{q_{1}q_{2}\over\mu}{{\bf r}\over r^{3}}+
{2\over 3c^{3}}\biggl({q_{1}\over m_{1}}-{q_{2}\over m_{2}}\biggr)
\frac{d}{dt}(q_{1}{\bf{a}}_{1}+q_{2}{\bf{a}}_{2}) \,,
\end {equation}
where $\mu = m_1 m_2 /(m_1 + m_2)$ is the reduced mass.
From equations (\ref{eqmov-p1}), (\ref{eqmov-p2}) and
(\ref{eqmovrel1}), it is easily shown that, keeping only the
lowest order ($c^{0}$) terms,
\begin {equation}
q_{1}{\bf{a}}_{1}+q_{2}{\bf{a}}_{2} =
\mu \left({q_{1}\over m_{1}}-{q_{2}\over m_{2}}\right)
{d^{2}{\bf r}\over dt^{2}} \,.
\end {equation}
Substituting this result in Eq.~(\ref{eqmovrel1}) we obtain
\begin {equation}
\label {eqmovfinal}
{d^{2}{\bf r}\over dt^{2}}=
{q_{1}q_{2}\over\mu}{{\bf r}\over r^{3}}+
\frac{2 \tilde q^2}{3 \mu c^3}\frac{d^3{\bf r}}{dt^3}\,,
\end {equation}
where
\begin {equation}
\label {deftildeq}
\tilde {q}=\mu\biggl({q_{1}\over m_{1}}-{q_{2}\over m_{2}}\biggr)\ .
\end {equation}

In the fixed target limit, $m_2 \to \infty$, Eq.~(\ref{eqmovfinal}) 
becomes the nonrelativistic Lorentz-Abraham equation of motion. 
It is interesting to see that two-body recoil effects appear in 
Eq.~(\ref{eqmovfinal}) not only through the reduced mass $\mu$, but 
also via the effective charge $\tilde q$.
In particular, if $q_{1}/m_{1}=q_{2}/m_{2}$ we have $\tilde {q}=0$, 
and there is no radiation reaction even though both particles are
accelerating. This is related to the fact that, in this case, there is no 
electric dipole radiation from the system.


\section{Radiative correction to Rutherford scattering}

In the absence of perturbations, Rutherford scattering conserves the total
energy $E = \frac{1}{2} \mu v^2 + q_1 q_2 / r$, the angular momentum 
${\bf L} = \mu {\bf r}\times{\bf v}$, and the Runge-Lenz vector \cite{LandauMech}
\begin{equation}
{\bf A} = {\bf v}\times{\bf L} + q_1q_2 \, {\bf\hat r} \,.
\end{equation}
Here, ${\bf v}=d{\bf r}/dt$ is the relative velocity and 
${\bf\hat r}={\bf r}/r$ is the radial unit vector.
These conserved quantities are not independent: it is
easily seen that ${\bf A}\cdot{\bf L}=0$ and
\begin{equation}
\label {ARL}
A^2 = 2 E L^2 / \mu + (q_1q_2)^2 = (v_0 L)^2 + (q_1 q_2)^2 \,,
\end{equation}
where $v_0$ is the initial (asymptotic) velocity.
Taking the scalar product ${\bf r}\cdot{\bf A}$, one finds
the Rutherford scattering orbit
\begin {equation}
\label {r-phi}
r(\varphi)=\frac{L^{2}/\mu}{A\cos\varphi -q_{1}q_{2}}\,,
\end {equation}
where $\varphi$ is the angle between ${\bf r}$
and ${\bf A}$. During the collision, $\varphi$ changes from
$-\varphi_0$ to $\varphi_0$, where 
\begin {equation}
\label{phi0}
\varphi_0 = \cos^{-1}(q_1q_2/A) = \tan^{-1}(v_0 L/q_1q_2) \,.
\end {equation}
The scattering angle is $\theta = \pi - 2 \varphi_0$, and from 
Eq.~(\ref{phi0}) we obtain the Rutherford deflection function
\begin {equation}
\theta(L)= 2 \tan^{-1}(q_1q_2/v_0 L) \,.
\end {equation}
Note that for charges of the same sign the scattering angle is positive,
and for opposite charges $\theta$ is negative (we take $L$ and
$v_0$ as always positive).

\begin{figure}
\begin{center}
\includegraphics[width=7cm]{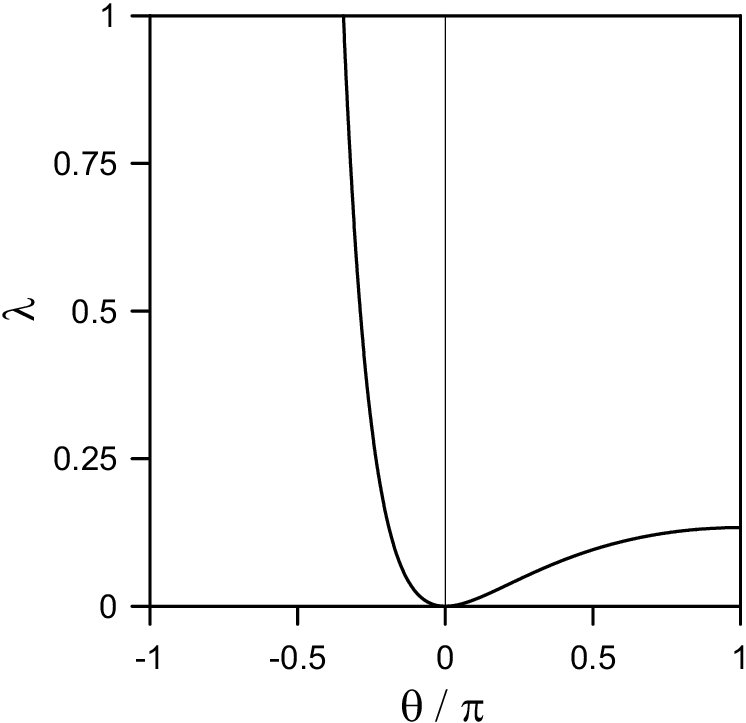}
\end{center}
\caption{Angular dependence of the radiative correction 
to Rutherford's deflection function.
Positive (negative) angles correspond to the scattering 
of like (unlike) charges.}
\label{lambda-theta}
\end{figure}

When radiation damping is considered, $E$, ${\bf L}$ and ${\bf A}$ are 
no longer conserved. 
In particular, from Eq.~(\ref{eqmovfinal}) we can show that the 
Runge-Lenz vector changes at the rate
\begin{equation}
\frac{d{\bf A}}{dt} = 
\frac{2 \tilde q^2}{3 c^3} \left[ \frac{1}{\mu} \frac{d^3{\bf r}}{dt^3}\times{\bf L} +
{\bf v}\times\left({\bf r}\times\frac{d^3{\bf r}}{dt^3}\right) \right] \,.
\end{equation}
The total change of ${\bf A}$ during the collision is then
\begin{equation}
\label{dA}
\delta {\bf A} = \frac{2 \tilde q^2}{3 c^3}\int_{-\infty}^{\infty} dt
\left[ \frac{1}{\mu} \frac{d^3{\bf r}}{dt^3}\times{\bf L} +
{\bf v}\times\left({\bf r}\times\frac{d^3{\bf r}}{dt^3}\right) \right]\,.
\end{equation}
The change of the Runge-Lenz vector is of order $c^{-3}$. 
Keeping the same order of approximation, we can substitute in the integrand of 
Eq.~(\ref{dA}) the results of unperturbed Rutherford scattering. 
We obtain
\begin {equation}
\delta {\bf A} = \frac{2 \tilde q^2}{3 c^3} \frac{q_1q_2}{\mu^2}
\int_{-\infty}^{\infty}dt \frac{{\bf A} - q_1q_2 \, {\bf \hat r}}{r^3} \,,
\end {equation}
which is further simplified by a change of variable from time
$t$ to angle $\varphi$. Still working to order $c^{-3}$, we have 
\begin {equation}
dt = \frac{\mu r^2}{L} d\varphi \,,
\end {equation}
and 
\begin {equation}
\delta {\bf A} = \frac{2 \tilde q^2}{3 c^3} \frac{q_1q_2}{\mu L} 
  \int_{-\varphi_0}^{\varphi_0}d\varphi \frac{{\bf A} - q_1q_2 \, {\bf \hat r}}{r} \,.
\end {equation}
Substituting $r(\varphi)$ from Eq.~(\ref{r-phi}), the above integral reduces to
\begin {equation}
\delta {\bf A} = \frac{2 \tilde q^2}{3 c^3} \frac{q_1q_2}{{L}^3 A} {\bf A}
\int_{-\varphi_0}^{\varphi_0}d\varphi 
\left(A \cos\varphi - q_1q_2 \right) \left(A - q_1q_2 \cos\varphi \right)
\,,
\end {equation}
which is easily calculated. Using Eq.~(\ref{phi0}), the result is written as
\begin {equation}
\delta {\bf A} = \frac{2 \tilde q^2}{3 c^3} \frac{q_1q_2 v_0}{L^2}
\left[ 2 + \frac{1}{1+(v_0 L/q_1q_2)^2} 
   - 3\frac{q_1q_2}{v_0 L} \tan^{-1}(v_0 L/q_1q_2) \right] {\bf A} \,.
\end {equation}
According to Eq.~(\ref{phi0}), the change in the Runge-Lenz vector modifies 
the asymptotic angle $\varphi_0$ by 
\begin{equation}
\delta\varphi_0 = \frac{q_1q_2}{v_0 L}\,\frac{\delta A}{A} \,,
\end{equation}
and the scattering angle $\theta$ by (see Ref.~\cite{Aguiar})
\begin{equation}
\delta \theta = - \delta \varphi_0 \,.
\end{equation}
The deflection function is then given as 
\begin{equation}
\theta(L) = 2 \tan^{-1}(q_1q_2/v_0 L) + \delta\theta(L)
\end{equation}
where the first term is the Rutherford relation, and the radiation damping 
correction is
\begin{equation}
\delta \theta (L) = - \frac{2 \tilde q^2}{3 c^3} \frac{(q_1q_2)^2}{L^3}
        \left[ 2 + \frac{1}{1+(v_0 L/q_1q_2)^2} 
      - 3\frac{q_1q_2}{v_0 L} \tan^{-1}(v_0 L/q_1q_2) \right] \,.
\end{equation}
From these equations we can also obtain $L(\theta)$. To order $c^{-3}$,
the result is
\begin{equation}
\label{Lc3}
L(\theta) = \frac{q_1q_2}{v_0} \cot (\theta/2) \left[ 1 + 
\frac{\tilde{q}^2}{q_1q_2} \left(\frac{v_0}{c}\right)^3 \lambda(\theta) \right]  \,,
\end{equation}
where
\begin{equation}
\label{lambda}
\lambda(\theta) = \frac{1}{6} \frac{\sin^3(\theta/2)}{\cos^5(\theta/2)}
  \left[ (5-\cos\theta) \cot(\theta/2) - 3(\pi-\theta) \right] \,.
\end{equation}
A plot of $\lambda(\theta)$ is shown in Fig.~\ref{lambda-theta}. As already 
mentioned, positive angles are reached by like-sign charges, and negative angles 
by oppositely charged particles. 
We see that the radiative correction is limited if the Coulomb force is repulsive, 
and is strongly divergent for backscattering ($\theta \rightarrow - \pi$) in an attractive
Coulomb field. 

\begin{figure}
\begin{center}
\includegraphics[width=7cm]{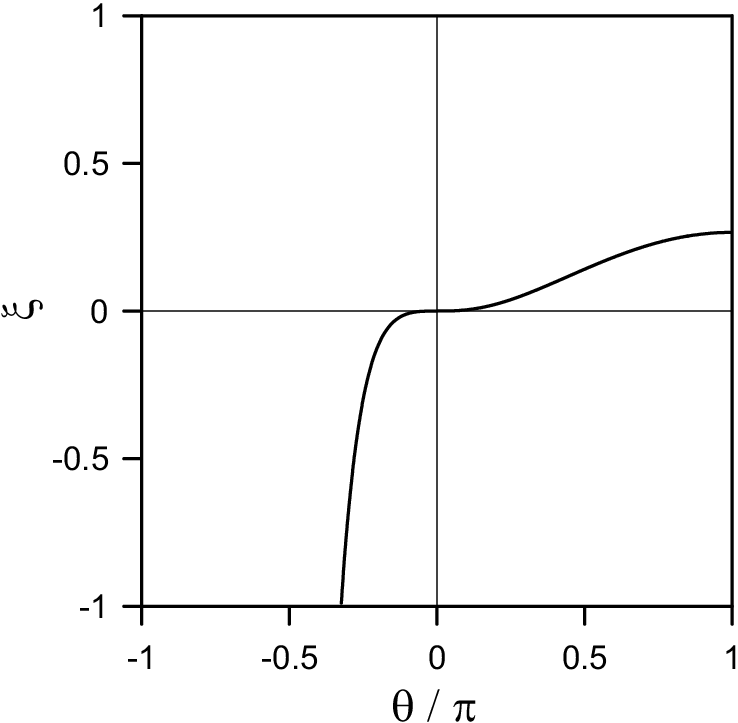}
\end{center}
\caption{Angular dependence of the radiative correction to 
the Rutherford cross section. Positive/negative angles are
the same as in Fig.~\ref{lambda-theta}.}
\label{xi-teta}
\end{figure}

The scattering cross section can be calculated from the deflection
as
\begin{equation}
\frac{d\sigma}{d\Omega} = 
\frac{1}{p^2} \left| \frac{L}{\sin \theta} \frac{dL}{d\theta} \right| \,,
\end{equation}
where $p=\mu v_0$ is the initial momentum.
With Eqs.~(\ref{Lc3}) and (\ref{lambda}) we get
\begin{equation}
\label{dsigma3}
\frac{d\sigma}{d\Omega} = \frac{d\sigma_{R}}{d\Omega} \left[ 1 + 
\frac{\tilde{q}^2}{q_1q_2} \left(\frac{v_0}{c}\right)^3 \xi(\theta) \right]\,,
\end{equation}
where 
\begin{equation}
\label{rutherford}
\frac{d\sigma_R}{d\Omega} = 
   \left( \frac{q_1q_2}{2 \mu {v_0}^2} \right)^2 \frac{1}{\sin^4(\theta/2)}
\end{equation}
is the nonrelativistic Rutherford cross section, and
\begin{equation}
\label{xi-theta}
\xi(\theta)= \frac{1}{2} \frac{\sin^3(\theta/2)}{\cos^5(\theta/2)} 
\left[ (\pi-\theta) (2-\cos\theta) - 3 \sin\theta \right]\,.
\end{equation}
The function $\xi(\theta)$ is shown in Fig.~\ref{xi-teta}. At large angles, close
to backscattering, $\xi(\theta)$ has the limits
\begin{eqnarray}
\xi(\theta)&\sim& \frac{4}{15} - \frac{2}{35} (\theta-\pi)^2 + \dots  \qquad 
(\theta\rightarrow\pi)
\\
\xi(\theta)&\sim& -96\pi (\theta+\pi)^{-5} + \dots  \qquad (\theta\rightarrow-\pi) 
\end{eqnarray}

\begin{figure}
\begin{center}
\includegraphics[width=7cm]{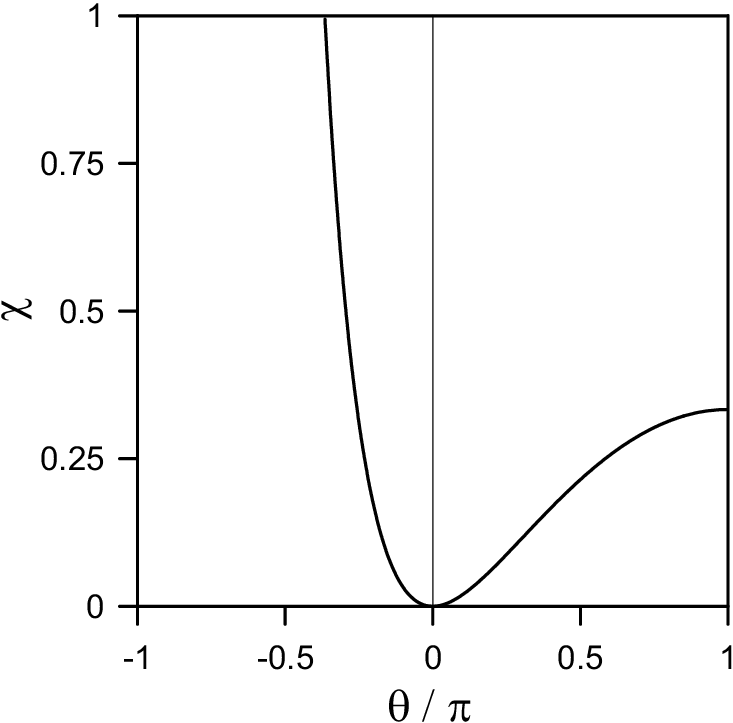}
\end{center}
\caption{Angular dependence of the change in angular 
momentum. Positive/negative angles are
the same as in Fig.~\ref{lambda-theta}.}
\label{chi-teta}
\end{figure}

The angular momentum loss (or gain) can be calculated with similar methods. 
With radiation damping, the time derivative of ${\bf L}$ is given by
\begin{equation}
\label{dLdt}
\frac{d{\bf L}}{dt} = 
\frac{2 \tilde q^2}{3 c^3} \, {\bf r} \times \frac{d^3{\bf r}}{dt^3} \, ,
\end{equation}
which, integrated on the unperturbed Rutherford trajectory,
gives the total change of angular momentum in the 
scattering process,
\begin {equation}
\label {deltaL}
\delta{\bf L} = \frac{4{\tilde q}^2}{3c^3} \frac{q_1q_2 v_0}{L^2} 
\left[1-\frac{q_1 q_2}{v_0 L} \arctan\left(\frac{v_0 L}{q_1q_2}\right) \right] {\bf L} \,.
\end {equation}
At a given scattering angle, the angular momentum change is
\begin {equation}
\label {deltaL-theta}
\delta{\bf L} = 
\frac{4}{3} \frac{{\tilde q}^2}{q_1 q_2} \left(\frac{v_0}{c}\right)^3 
\chi(\theta) \, {\bf L} 
\end {equation}
where
\begin {equation}
\label {chi-theta}
\chi(\theta) = 
\tan^2(\theta/2) \left[1-\frac{\pi-\theta}{2} \tan(\theta/2) \right] \,.
\end {equation}
This function is shown in Fig.~\ref{chi-teta}

The energy loss is readily calculated by differentiating Eq.~(\ref{ARL}),
\begin{equation}
\frac{\delta E}{E} = \frac{2}{(v_0 L)^2} {\bf A}\cdot{\delta{\bf A}}
    - \frac{2}{L^2}{\bf L}\cdot{\delta{\bf L}}\, .
\end{equation}
Inserting the expressions for $\delta{\bf A}$ and $\delta{\bf L}$ we obtain
\begin{equation}
\frac{\delta E}{E} = \frac{4{\tilde q}^2}{3c^3} \frac{(q_1q_2)^2}{L^3} 
\left\{ 3 \frac{q_1 q_2}{v_0 L} - 
\left[1+3 \left(\frac{q_1 q_2}{v_0 L}\right)^2\right] 
\arctan\left(\frac{v_0 L}{q_1q_2}\right) \right\} \,,
\end{equation}
or, in terms of the scattering angle,
\begin{equation}
\frac{\delta E}{E} = - \frac{4}{3} 
\frac{{\tilde q}^2}{q_1 q_2} \left(\frac{v_0}{c}\right)^3 \xi(\theta) \,
\end{equation}
where $\xi(\theta)$ is the same function given in Eq.~(\ref{xi-theta}) 
and shown in Fig.~\ref{xi-teta}.


\section{Final comments}

Our discussion of radiation damping corrections to Rutherford scattering 
ignored relativistic effects like retardation, magnetic forces, and the 
mass-velocity dependence. These effects give contributions of order 
$c^{-2}$ to the deflection function and cross section (see Ref.~\cite{Aguiar}), 
and are generally more important than the $c^{-3}$ radiative corrections we
have obtained. 
They were not considered here because, as already mentioned, this would not
change our results: a $c^{-2}$ correction to the nonrelativistic Rutherford 
trajectory only adds $c^{-5}$ terms to our perturbative calculation of
radiation damping. 
We can easily write the complete (up to $c^{-3}$) expansion of the deflection 
function and scattering cross section by putting together the results of 
Ref.~\cite{Aguiar} and the present paper. 
For example, the differential cross section to order $c^{-3}$ reads
\begin{equation}
\label{dsig-2-3}
\frac{d\sigma}{d\Omega} = \frac{d\sigma_{R}}{d\Omega} 
\left[ 1 - \left(\frac{v_0}{c}\right)^2 h(\theta) \right]
\left[ 1 + 5 \frac{\mu}{M} \left(\frac{v_0}{c}\right)^2 \right]
\left[ 1 + \frac{\tilde{q}^2}{q_1q_2} \left(\frac{v_0}{c}\right)^3 \xi(\theta) \right]\,,
\end{equation}
where 
\begin{equation}
h(\theta) = \frac{1}{2} \tan^2(\theta/2) 
\left[ 1 + (\pi -\theta) \cot\theta \right] +1
\end{equation}
and $M=m_1+m_2$. 
As discussed in \cite{Aguiar}, the first corrective term accounts for the 
variation of mass with velocity, and the second includes magnetic and 
retardation effects. The last one is the radiative correction calculated 
in the previous section. 
It is interesting to note that magnetic and retardation effects simply 
renormalize the cross section by an angle independent factor.

\begin{figure}
\begin{center}
\includegraphics[width=7cm]{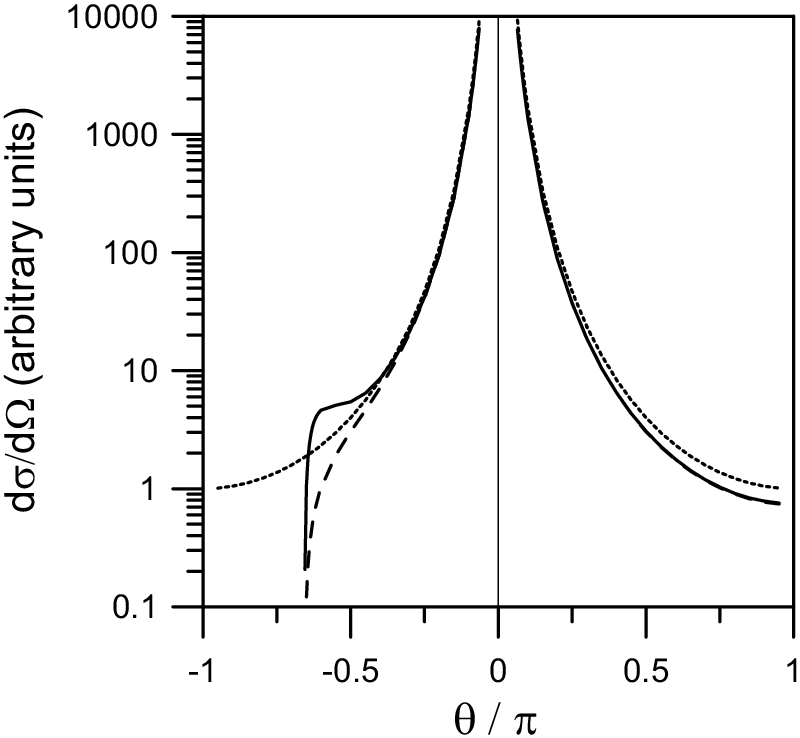}
\end{center}
\caption{Rutherford scattering to order $c^{-3}$. The projectile 
velocity is $0.4\, c$, and the target has infinite mass. 
The two electric charges are of the same magnitude, like (unlike) signs 
corresponding to positive (negative) scattering angles. 
The nonrelativistic Rutherford cross section is given by the dotted lines. 
The dashed lines incorporate $c^{-2}$ corrections, and the solid lines  
include the $c^{-3}$ radiation damping effects.}
\label{fig-sigma}
\end{figure}

In Fig.~\ref{fig-sigma} we show the differential cross section for the scattering
of a charged particle with $v_0=0.4\,c$ on a fixed target, of equal ($\theta>0$) 
or opposite ($\theta<0$) charge. The dotted lines give the nonrelativistic
cross section, and the dashed ones show the effect of the $c^{-2}$ relativistic mass 
correction (retardation and magnetic forces do not show up on a fixed target).
The solid lines bring in the radiation damping effect, as given in Eq.~(\ref{dsig-2-3}).
We see in Fig.~\ref{fig-sigma} that radiation damping has a very small effect
when the charges repel each other. But for an attractive Coulomb force the 
radiative correction is quite important (as also seen in Fig.~\ref{xi-teta}), 
creating a plateau-like structure in the angular distribution. 
Even though our perturbative results are not reliable for large corrections, 
such structure is very similar to what is found in ``exact'' numerical 
calculations \cite{Huschilt}.

A final point we wish to comment on is why our results are not plagued by 
runaway solutions. The reason is that the Runge-Lenz based perturbative 
calculation presented here follows essentially a ``reduction of order'' 
approach, such as described in Refs.~\cite{Landau,Wald}. This effectively
eliminates the additional degrees of freedom introduced in the equations of 
motion by the time derivative of acceleration, yielding only physically
acceptable solutions.


\end {document}